\documentclass[twocolumn,tighten,times]{aastex62}
\usepackage{color}
\usepackage{multirow}
\usepackage{chngcntr}
\usepackage{mathtools}
\usepackage{amsmath}
\usepackage{amsfonts}
\usepackage{amssymb}
\usepackage[utf8]{inputenc}  % 使用标准的 utf8 编码
\usepackage{graphicx}        % 确保这里是打开的
\usepackage{perpage}         % 在脚注中使用
\usepackage{bm}
\MakePerPage{footnote}

\graphicspath{{./}{figures/}}

\makeatletter

\newcommand{\Rmnum}[1]{\expandafter\@slowromancap\romannumeral #1@}
\makeatother

\shorttitle{Halo spin vs. HI fraction}
\shortauthors{Liu et al.}

\begin{document}

\title{Strong Correlation between Galactic HI-to-stellar Mass Ratio And Halo Spin Explored by HI-rich Galaxies}

\correspondingauthor{Yu Rong}
\email{rongyua@ustc.edu.cn}

\author{Shihong Liu}
\affiliation{Department of Astronomy, University of Science and Technology of China, Hefei, Anhui 230026, China}
\affiliation{School of Astronomy and Space Sciences, University of Science and Technology of China, Hefei 230026, Anhui, China}

\author{Yu Rong$^*$}
\affiliation{Department of Astronomy, University of Science and Technology of China, Hefei, Anhui 230026, China}
\affiliation{School of Astronomy and Space Sciences, University of Science and Technology of China, Hefei 230026, Anhui, China}

\author{Zichen Hua}
\affiliation{Department of Astronomy, University of Science and Technology of China, Hefei, Anhui 230026, China}
\affiliation{School of Astronomy and Space Sciences, University of Science and Technology of China, Hefei 230026, Anhui, China}

\author{Huijie Hu}
\affiliation{University of Chinese Academy of Sciences, Beijing 100049, China}



\begin{abstract}

	Using a semi-analytic approach, we estimate halo spins for a large sample of HI-rich galaxies from the Arecibo Legacy Fast Alfa Survey and examine the correlation between HI mass fractions and halo spins. Our analysis reveals a strong correlation between halo spin and the HI-to-stellar mass ratio in both low-mass and massive galaxy samples. This finding suggests a universal formation scenario: higher halo spin reduces angular momentum loss and gas condensation, leading to lower star formation rates and weaker feedback, which in turn helps retain gas within dark matter halos.

\end{abstract}


\section{Introduction}           %% first-level sections will be auto-capitalized
\label{sec:1}

In the Lambda cold dark matter framework, the characteristics of galaxies are determined by the properties of their host dark matter halos. Factors such as gas accretion mode, gas temperature, angular momentum, and star formation efficiency within a galaxy are influenced by the characteristics of the dark matter halos \citep[e.g.,][]{Rubin10,Rubin15,Lehner14,Fardal01, Keres05,vandeVoort12,Nelson13,Noguchi23,Guo11,Yang12,Behroozi10,Girelli20,Silk77}. While various halo properties like halo mass and concentration play a role in empirical galaxy formation models, the influence of halo spin on galaxy features may be particularly significant but is not yet fully understood. It is believed that the spin of parent dark matter halos affects, and possibly determines, the distribution of stars \citep{Mo98,vandenBosch98,Diemand05,Desmond17}. The prevailing theoretical expectation is that gas in halos with higher spin will exhibit faster rotation, resulting in a more dispersed distribution and lower stellar density during star formation.

Previous hydrodynamical simulations (e.g., \citealt{Kim13,Jiang192}) and semi-analytic galaxy formation models (e.g., \citealt{Mo98,vandenBosch98}) have shown that halo spin strongly influences the size and density of stellar matter distribution, especially in massive late-type galaxies. However, recent studies based on high-resolution simulations suggest that the role of halo spin in low-mass galaxies is still debated: for most typical dwarf galaxies, their stellar distributions may be unaffected by or only weakly correlated with halo spin \citep{Yang23,DiCintio19}, while for specific dwarf galaxy populations like ultra-diffuse galaxies (\citealt{Rong17a,Amorisco16,Liao19,Benavides23}), halo spin could significantly impact their stellar distributions.

In addition to the influence of halo spin on the distribution of stars \citep{Rong24c}, the HI fraction within halos may also be impacted by spin \citep[e.g.,][]{Obreschkow16,Rong24a}. Cosmological simulations \citep[e.g.,][]{
Lagos16,Nelson15} suggest that the HI fraction provides insights into a galaxy's history and plays a significant role in its future growth and morphology. For example, \cite{Rong24a} discovered that excessive halo spin could impede gas from shedding angular momentum, making it challenging for the gas to accumulate at the halo's core and form stars. This often results in a decreased rate of star formation, weakening stellar feedback mechanisms (such as supernova feedback) and inadequately expelling gas from the halo, ultimately leading to an increased gas fraction within galaxies.

While the impact of halo spin on HI fractions and HI-to-stellar mass ratios in galaxies remains inconclusive, especially in observational studies where determining halo spin poses a challenge. Measurements of halo spin in galaxies have primarily focused on a limited number of samples with resolved HI observations \citep[e.g.,][]{Hunter12,Oh15}, which are insufficient for statistically analyzing the relationship between halo spin and stellar distribution in galaxies. Large-sample integral field unit (IFU) observations like MaNGA mainly concentrate on regions within halos dominated by stellar matter, making accurate estimation of halo spin challenging with IFU data (e.g., \citealt{Wang20,Rong18,Cappellari06,Cappellari13}). Furthermore, due to constraints in observation time and telescope sensitivity, IFU observations tend to target galaxies with higher surface brightness and stellar masses, leading to sample selection bias. Even with accurate measurements of halo spin parameters, studying the correlation between halo spin and HI-to-stellar mass ratios within galaxies accurately and without bias remains challenging.

 Alternatively, HI surveys conducted with single-dish telescopes such as the comprehensive Arecibo Legacy Fast Alfa Survey \citep[ALFALFA;][]{Giovanelli05,Haynes18} and the undergoing FAST All Sky HI survey \citep[FASHI;][]{Zhu23}, provide valuable opportunities to obtain HI spectra from numerous galaxies. These surveys offer essential dynamical information on galaxies, enabling the estimation of spin parameters across a large galaxy sample and facilitating comparisons of halo spins in galaxies with diverse mass distributions. Galaxy samples selected by HI surveys have an unbiased stellar mass distribution, making them suitable for studying the correlation between halo spin and HI-to-stellar mass ratios.

In this study, we first utilize a semi-analytic approach to estimate halo spin for each HI-bearing galaxy cataloged in ALFALFA, and then investigate the potential correlation between halo spins and HI-to-stellar mass ratios of galaxies. Section~\ref{sec:2} presents the sample data and outlines the methodology for estimating halo spin. Section~\ref{sec:3} provides a statistical analysis of the dependence of HI-to-stellar mass ratios of galaxies on their halo spins. Our conclusions are presented in section~\ref{sec:4}.

\section{Data}
\label{sec:2}

\subsection{Sample}

 The galaxy sample is sourced from the ALFALFA extragalactic HI survey, covering approximately 6,600 deg$^2$ at high Galactic latitudes. The comprehensive ALFALFA catalog \citep[$\alpha.$100;][]{Haynes18}, contains around 31,500 sources with radial velocities below 18,000 km s$^{-1}$, providing properties for each source, including HI spectrum signal-to-noise ratio (SNR), cosmological distance, HI line width at 50\% peak ($W_{50}$) corrected for instrumental broadening, and HI mass ($M_{\rm{HI}}$).

\subsection{HI-to-stellar mass ratio}

ALFALFA galaxies are cross-matched with SDSS data \citep{Alam15}. Following \cite{Durbala20}, stellar masses ($M_{\star}$) of ALFALFA galaxies with optical counterparts are estimated using three methods: UV-optical-infrared SED fitting, SDSS $g-i$ color, and infrared $W_2$ magnitude. This study primarily uses stellar mass from SED fitting, and for galaxies lacking UV or infrared data for SED fitting, $g-i$ color is used to estimate stellar mass. Discrepancies among methods are considered negligible.

The HI-to-stellar mass ratio, $\eta$, is then calculated as $\eta = \log M_{\rm{HI}} - \log M_{\star}$, where both $M_{\star}$ and $M_{\rm{HI}}$ are in units of $M_{\odot}$.

\subsection{Rotation velocity and halo spin}

The rotation velocity is calculated as $V_{\rm{rot}}=W_{50}/2/\sin\phi$, where $\phi$ represents the inclination of the HI disk. When resolved HI data is unavailable, we estimate the HI disk inclination $\phi$ using the optical axis ratio $b/a$ from \cite{Durbala20} as $\sin\phi=\sqrt{(1-(b/a)^2)/(1-q_0^2)}$, where $q_0\sim 0.2$ \citep{Tully09,Giovanelli97,Li21} for massive galaxies, and $q_0\sim 0.4$ (\citealt{Rong24}) for low-mass galaxies with $M_{\star}<10^{9.5}\ M_{\odot}$. if $b/a \leq q_0$, we set $\phi=90^{\circ}$.

To improve rotation velocity accuracy, we exclude galaxies with inclinations $\phi<50^{\circ}$ and those with low HI SNRs (SNR$<10$) due to high velocity uncertainties.

Some HI-bearing galaxies may exhibit velocity dispersion-dominated kinematics rather than regular rotation, often characterized by `single-horned' HI line profiles \citep{ElBadry18}, which complicates accurate rotation and halo spin measurements. Following \cite{Hua24}, we distinguish single-horned from double-horned HI spectra using the kurtosis parameter $k_4$, classifying spectra as single-horned if $k_4>-1.0$. Our analysis focuses exclusively on isolated galaxies with double-horned profiles, excluding potentially dispersion-dominated systems.

For rotation-dominated galaxies with high HI SNR and inclinations, we estimate halo spin following \citealt{Hernandez07}:
\begin{equation}
\lambda_{\rm{h}}\simeq 21.8 \frac{R_{\rm{HI,d}}/{\rm kpc}}{(V_{\rm{rot}}/{\rm km s^{-1}})^{3/2}}.
	\label{sam_HI} \end{equation}
Here, the galaxy’s dark matter halo is assumed to follow an isothermal sphere model, with baryonic gravitational effects neglected.
The HI disk scale length, $R_{\rm{HI,d}}$, is derived assuming a thin, rotationally supported gas disk with an exponential surface density profile \citep{Mo98}:
\begin{equation}
	\Sigma_{\rm{HI}}(R)=\Sigma_{{\rm{HI}},0} {\rm{exp}}(-R/R_{{\rm{HI,d}}}),
\end{equation}
where $\Sigma_{{\rm{HI}},0}$ is the central surface density. Total HI mas $M_{\rm{HI}}$ relates to $R_{{\rm{HI,d}}}$ as:
\begin{equation} \int_0^\infty \Sigma_{\rm{HI}}(R)2\pi R{\rm d}R = 2 \pi \Sigma_{{\rm{HI}},0} R_{{\rm{HI,d}}}^2 = M_{\rm{HI}}\label{HIeq_mass}. \end{equation} 
We define the HI radius $r_{\rm{HI}}$ as the radius where HI surface density reaches $1\ \rm M_{\odot}\rm{pc^{-2}}$. Empirical studies provide $r_{\rm{HI}}$ in terms of HI mass $M_{\rm{HI}}$ as $\log r_{\rm{HI}}=0.51\log M_{\rm{HI}}-3.59$ \citep{Wang16,Gault21}. Thus, at $r_{\rm{HI}}$:
\begin{equation} \Sigma_{{\rm{HI}},0} {\rm{exp}}(-r_{\rm{HI}}/R_{{\rm{HI,d}}})=1\ \rm  M_{\odot}\rm{pc^{-2}}. \label{HIeq_3} \end{equation} 
Using equations above, we compute $R_{{\rm{HI,d}}}$ for each galaxy, allowing halo spin estimation.

\subsection{Environment control}

Environmental factors exert a profound influence on the HI fractions within galaxy halos \citep{Yozin15, Davies19, Kawata08}. Over-dense environments, such as clusters, groups or even close pairs \citep{Patton11,
Robotham14,Davies16a} can significantly impair or entirely inhibit the inflow of gas essential for sustained star formation, thereby triggering quenching events \citep{Peng10,Schaefer17}. Observational studies have established that the gas fractions, star formation rates, and colors-indicators of star formation activity-of galaxies exhibit a correlation with local galaxy density within groups and clusters \citep{Peng12,Treyer18,Davies19,Forman79},
as well as with their positions relative to the center of these systems \citep{Wolf09,Wetzel12,Woo15,Barsanti18,Martinez08}. Furthermore, previous investigations, alongside our recent analyses, have indicated that halo spin is contingent upon environmental factors and the effects of tidal fields \citep{Hahn07a,Hahn07b,Wang17,Xue24,Hua24,Wang11}.
Consequently, any potential correlation between halo spin and the HI-to-stellar mass ratio may be obscured or entirely negated by environmental influences. Therefore, it is imperative that we account for environmental variables when examining the dependence of the ratio $\eta$ on halo spin. We employ the galaxy group and cluster catalog developed by \cite{Saulder16}, derived from SDSS DR12 \citep{Alam15} and the 2MASS Redshift Survey \citep{Huchra12}, which is constructed using the friends-of-friends algorithm and rigorously corrects for observational biases, such as the Malmquist bias and the `Fingers of God' effect. 

To eliminate environmental influences, we select only isolated galaxies, defined as those located beyond three times the virial radius from any galaxy group or cluster \citep{Guo20,Rong24a}. This threshold effectively excludes regions subject to tidal or ram-pressure stripping associated with dense environments \citep{Mamon04, Gill05, Oman13, Zinger18}.

%%%%%%%%%%%%%%%%%%%%%%%%%%%%%%%%%%%%%%

\section{Results}
\label{sec:3}

\cite{DiCintio19} identify a stellar mass $\sim10^9\ M_{\odot}$ as a transition point between feedback-dominated and angular momentum-dominated galaxy formation regimes. This suggests that for low-mass galaxies, HI-to-stellar mass ratios $\eta$ may be less dependent on halo spin. To investigate, we divide the sample into low-mass ($M_{\star}<10^9\ M_{\odot}$) and massive ($M_{\star}>10^9\ M_{\odot}$) subsamples, analyzing their respective $\eta$-$\lambda_{\rm{h}}$ relationships.

\begin{figure*}
   \centering
   \includegraphics[width=0.9\textwidth, angle=0]{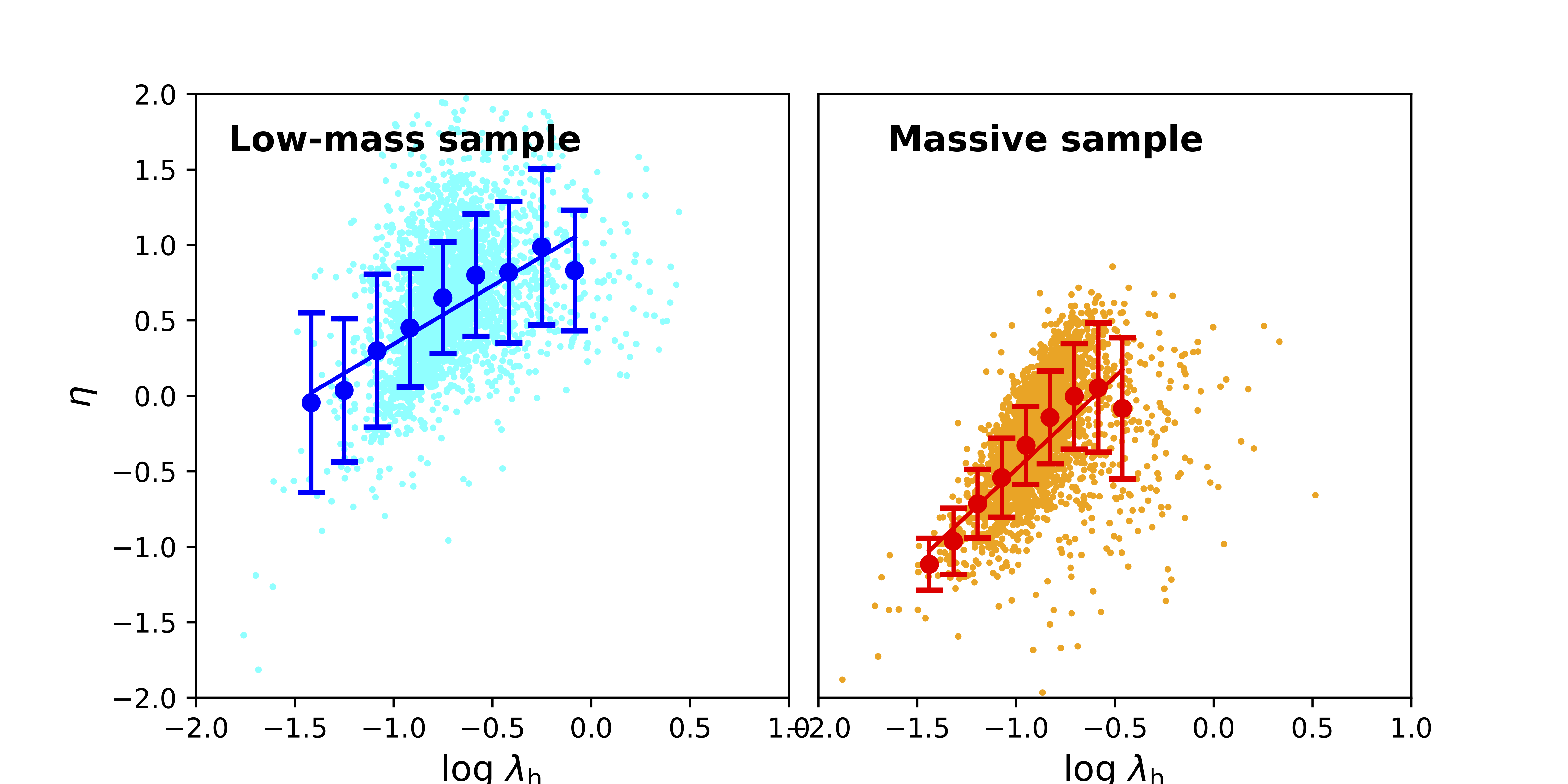}
   \caption{The $\eta$-$\lambda_{\rm{h}}$ relationships for low-mass (panel~a) and massive (panel~b) galaxies. Median $\eta$ values with $1\sigma$ error bars are shown in blue and red for bins in $\log \lambda_{\rm{h}}$. The best linear fitting results are highlighted by the corresponding lines.}
   \label{fig1}
   \end{figure*}

Figure~\ref{fig1} presents $\eta$ as a function of $\lambda_{\rm{h}}$ for low-mass (left panel) and massive (right panel) galaxies. Median $\eta$ values with $1\sigma$ error bars are shown in blue and red for bins in $\log \lambda_{\rm{h}}$. 
Our findings indicate that massive galaxies have generally lower HI fractions than low-mass galaxies; however, in both subsamples, $\eta$ increases significantly with halo spin. The correlation coefficients are 0.40 for low-mass and 0.50 for massive galaxies, demonstrating a strong $\eta$-$\lambda_{\rm{h}}$ correlation. Linear fits yield  $\eta\simeq (0.73\pm 0.37)\log \lambda_{\rm{h}}+(1.09\pm 0.30)$ for low-mass galaxies and $\eta\simeq (1.39\pm 0.31)\log \lambda_{\rm{h}}+(0.91\pm 0.36)$ for massive galaxies.

While our estimates of rotation velocities and halo spins are derived from optical inclinations, this approach may introduce uncertainty due to minor misalignments (${\rm d}\phi$) between optical and HI inclinations in certain galaxies \citep{Hunter12,Oh15,Nelson18,Nelson19}. Nonetheless, this discrepancy does not undermine the robust correlation observed between $\eta$ and $\lambda_{\rm{h}}$. Previous studies have demonstrated that the stellar and gas disks within galaxies may not be perfectly co-planar, frequently exhibiting a small inclination difference of ${\rm \delta}\phi < 20^\circ$ \citep{Starkenburg19,Guo20,Gault21}. However, while this inclination uncertainty may contribute to an increased scatter in the halo spins within our galaxy sample, it does not uniformly bias the measurements, resulting in either overestimation or underestimation of all spins.

%%%%%%%%%%%%%%%%%%%%%%%%%
\section{Conclusion}
\label{sec:4}

Using a semi-analytic approach, we estimate halo spins for the ALFALFA HI-bearing galaxy sample and investigate the potential dependence of HI-to-stellar mass ratios $\eta$ on halo spins $\lambda_{\rm{h}}$ for isolated low-mass and massive galaxies. We find a strong correlation between  $\eta$ and $\lambda_{\rm{h}}$ in both samples, suggesting that halo spin plays a central role in regulating gas cooling, star formation, and feedback in galaxies. This finding aligns with our previous results in \cite{Rong24a}.

As proposed by \cite{Rong24a}, a galaxy halo with increased spin would likely lead to high-spin gas accretion, where angular momentum loss is difficult. This high-spin gas resists infall towards the galactic core, slowing the condensation and cooling processes essential for central star formation \citep{Peng20}. Consequently, gas inflow and star formation proceed at a moderated, continuous rate, reducing supernova feedback intensity compared to low-spin halos. This mitigated feedback prevents significant HI gas loss from the halo, thereby sustaining a higher HI fraction.

Compared to \cite{Rong24a}, which centered on ultra-diffuse galaxies, we find that the $\eta$-$\lambda_{\rm{h}}$ correlation and formation mechanism appear applicable to galaxies across a broad mass range.
%\clearpage

%\begin{acknowledgments}
\acknowledgments

Y.R. acknowledges supports from the CAS Pioneer Hundred Talents Program (Category B), the NSFC grant 12273037, the USTC Research Funds of the Double First-Class Initiative. This work is supported by the China Manned Space Program with grant no. CMS-CSST-2025-A06 and CMS-CSST-2025-A08.

%\end{acknowledgments}






\end{document}